\documentstyle[12pt,fleqn]{article}
\def\double{\baselineskip 24pt \lineskip 10pt}

\def\be{\begin{equation}}
\def\ee{\end{equation}}
\def\bea{\begin{eqnarray}}
\def\eea{\end{eqnarray}}

\def\lz{{\lambda_0}}
\def\pz{{\phi_0}}

\def\vzp{{V'(\phi_0)}}
\def\fun#1#2{\lower3.6pt\vbox{\baselineskip0pt\lineskip.9pt
        \ialign{$\mathsurround=0pt#1\hfill##\hfil$\crcr#2\crcr\sim\crcr}}}
\begin{document}
\begin{titlepage}
\vspace*{-62pt}
\begin{flushright}
SUSSEX-AST 93/4-1 \\
FERMILAB--PUB--93/071--A\\
April 1993
\end{flushright}
\vspace{1.5cm}
\begin{center}
{\Large \bf Observing the Inflaton Potential}\\
\vspace{.6cm}
\normalsize
{Edmund J.\ Copeland,$^*$ Edward W.\ Kolb,$^{(\dag,\ddag)}$ \\
Andrew R.\ Liddle,$^{\S}$ and James E.\ Lidsey$^{\dag}$}\\
\vspace{24pt}

{\it
$^*$School of Mathematical and Physical Sciences,
University of Sussex,\\ Brighton BN1 9QH, U.\ K.\\
\vspace{18pt}
$^{\dag}$NASA/Fermilab Astrophysics Center,\\
Fermi National Accelerator Laboratory, Batavia, Illinois~~~60510\\
\vspace{18pt}
$^{\ddag}$Department of Astronomy and Astrophysics, The Enrico Fermi
Institute,\\The University of Chicago, Chicago, Illinois~~~60637\\
\vspace{18pt}
$^{\S}$Astronomy Centre, School of Mathematical and Physical Sciences,\\
University of Sussex, Brighton BN1 9QH, U.\ K.
}

\end{center}

\vspace*{12pt}

\double

\begin{quote}
\hspace*{2em}
We show how  observations of  the density perturbation (scalar) spectrum and
the gravitational wave (tensor) spectrum allow a reconstruction of the
potential responsible for cosmological inflation. A complete functional
reconstruction or a perturbative approximation about a single scale are
possible; the suitability of each approach depends on the data available.
Consistency equations between the scalar and tensor spectra  are derived, which
provide a powerful signal of inflation.
\vspace*{12pt}

PACS number(s): 98.80.--k, 98.80.Cq, 12.10.Dm

\vspace*{12pt}

\noindent
\small email: $^*$edmundjc@central.sussex.ac.uk;\qquad
$^{(\dag,\ddag)}$rocky@fnas01.fnal.gov;\\
\phantom{email: } $^{\S}$arl@starlink.sussex.ac.uk;
\qquad$^{\dag}$jim@fnas09.fnal.gov

\end{quote}

%%%%%%%%%%%%%%%%%%%%%%%%%%%%%%%%%%%%%%%%%%%%%%%%%%%%%%%%%%%%%%%%%%%%%%
\end{titlepage}
%%%%%%%%%%%%%%%%%%%%%%%%%%%%%%%%%%%%%%%%%%%%%%%%%%%%%%%%%%%%%%%%%%%%%%

\baselineskip=18pt

One of the most exciting aspects of the recent detection of large angle
microwave background anisotropies by COBE \cite{DMR} is the possibility that
part of the anisotropy observed is due to long wavelength gravitational wave
(tensor) modes instead of (scalar) density perturbations. In general the
influence of scalar and tensor modes on microwave background anisotropies
differs as a function of angular scale, and the use of measurements on
different scales may allow one to separate the anisotropies into their scalar
and tensor components. This has recently been considered by Crittenden {\it et
al} \cite{CBDES}.

This prospect is especially exciting for models of cosmological inflation,
proposed over a decade ago \cite{INF} as a possible resolution of a number of
otherwise puzzling aspects of the standard hot big bang cosmology \cite{KT}.
Inflation has long been known to predict that both scalar modes \cite{SCALREF}
and tensor modes \cite{STAR1,AW,STAR2} should exist on all astrophysically
relevant scales. Although the generic prediction from inflation has in the past
been advertised as a flat (Harrison--Zel'dovich) scalar spectrum and a tensor
spectrum of negligible amplitude, the rapid improvement of observational data
has led many researchers \cite{TENSORS} to emphasize recently the importance of
taking the detailed inflationary predictions seriously. Typically the
predictions from inflation are that the scalar spectrum possesses a scale {\em
dependence}, which is weak in many models but can be rather marked in others.
And though the amplitude of tensors may  typically be less than that of the
scalars, this does not necessarily imply that it is negligible.

This is problematic from a large scale structure viewpoint, since different
inflationary models offer a range of predictions and there is currently no
clear guidance from particle physics as to which inflationary models may be
suitable. The correct input one should make into a large scale structure model
is therefore unknown. However, from an inflationary viewpoint this is a
promising feature, as it raises the possibility that improved observations may
allow one to distinguish between inflationary models. The aim of this {\em
Letter} is to investigate the use of observations precisely to this end, by
deriving equations which allow one to proceed from a knowledge of the scalar
and/or tensor spectrum to a determination of the inflaton potential. As a very
useful by-product, we derive a consistency relation between the allowed
scale-dependences of the scalar and tensor modes. This is a powerful
discriminant for inflationary models in general, as it does not depend on a
specific choice of inflationary model.

Reconstruction of the inflaton potential in this manner was first considered
by Hodges and Blumenthal \cite{HB} (hereafter HB). We improve upon their
results in two important ways. Firstly, we consider both scalar and tensor
modes, whereas they restricted their study to the scalars alone. This is a
vital improvement, because, as HB acknowledged and we rederive, the scalars
alone are insufficient to uniquely determine the inflaton potential --- such a
reconstruction is possible only up to an undetermined constant, and as the
reconstruction equations are nonlinear this leads to functionally different
potentials giving rise to the same spectrum. The tensors (even just the tensor
amplitude at a single scale) provide just the extra information needed to lift
this degeneracy. Secondly, their analysis made explicit use of the so-called
{\em slow-roll} approximation. It is well known that this approximation breaks
down unless both the scalar spectrum is nearly flat and the tensor amplitude is
 negligible. We consider the inflation dynamics in full generality. However,
general expressions  for the perturbation spectra are not known, and one must
use slow-roll there. It is shown in an
accompanying paper \cite{CKLL} that this hybrid approach offers substantial
improvements over pure slow-roll results.

The equations of motion are most conveniently written in the $H(\phi)$
formalism \cite{LPLB}. An isotropic scalar field $\phi$ in a spatially flat
universe satisfies
\bea
\label{eq1a}
(H')^2-\frac{3}{2}\kappa^2 H^2 & = & -\frac{1}{2} \kappa^4 V(\phi)\\
\label{eq1b}
\kappa^2 \dot{\phi} & = & -2H' ,
\eea
provided that $\dot{\phi}$ does not pass through zero, where overdots are time
derivatives,  primes are derivatives with respect to  $\phi$, and $\kappa =
8\pi G = 8\pi/m_{Pl}^2$. The usual slow-roll approximation amounts to
neglecting  the first term in  Eq.\ (\ref{eq1a}) and its $\phi$-derivative.

The amplitudes of the scalar and tensor modes may be written using the
standard expressions as
\bea
\label{eq2a}
A_S(\phi) & = & \frac{\sqrt{2}\kappa^2}{8\pi^{3/2}} \,
\frac{H^2(\phi)}{|H'(\phi)|}\\
\label{eq2b}
A_G(\phi) & = & \frac{\kappa}{4\pi^{3/2}}\, H(\phi),
\eea
respectively. $A_S$ is equivalent to $P^{1/2}(k)/3\sqrt{2\pi}$ in HB, to
$\delta_H$ of Ref. \cite{LL2}, and for a flat spectrum equal to
$4\pi\epsilon_H$ of Ref. \cite{AW}. $A_G^2$ is equivalent to ${\cal P}_g/32\pi$
of Ref. \cite{LL2}. One  immediately notes that
\be
\label{eq10}
\frac{A_G}{A_S} = \frac{\sqrt{2}}{\kappa} \, \frac{|H'|}{H} =
\frac{\sqrt{2}}{\kappa} \left| \frac{{\rm d} \ln A_G}{{\rm d} \phi}
\right| ,
\ee
so the inflationary condition $\ddot{a} > 0$  implies $A_G <  A_S$. However,
the relative contribution of tensors to scalars for large angle
microwave background aniso\-tropies is given roughly (for sufficiently flat
spectra) by the ratio $25 A_G^2/2 A_S^2$ \cite{CKLL}, so it is possible for the
tensor contribution to dominate the anisotropy.

The spectra are quoted above as functions of $\phi$---that is, we are given
the amplitude when the scalar field takes a particular value. To compare
with observations we must relate $\phi$ to a given cosmological scale
$\lambda$. This is achieved by utilizing the formula
\be
\label{eq:NNN}
N(\phi) \equiv \int_{t_e}^{t} H(t') dt' =
- \frac{\kappa^2}{2} \int_{\phi }^{\phi_e} \frac{H(\phi')}{H'(\phi')}
d\phi',
\ee
which gives the number of $e$-foldings between a scalar field value $\phi$ and
the end of inflation at $\phi=\phi_e$. Each length scale $\lambda$ is
associated with a unique value of $\phi$ when that scale crossed the Hubble
radius during inflation, indicated by writing $\lambda(\phi)$. That value of
$\phi$ is also associated with a value $a(\phi)$ of the scale factor. We can
make use of Eq.\ (\ref{eq:NNN}) to relate $a(\phi)$ to the value of the scale
factor at the end of inflation, $a_e$: $a(\phi)=a_e\exp[-N(\phi)]$, which
allows us to express $\lambda(\phi)$ as
\be
\label{eq8}
\lambda(\phi)=\frac{\exp[N(\phi)]}{H(\phi)} \, \frac{a_0}{a_e}.
\ee
Differentiating Eq.\ (\ref{eq8}) with respect to ${\phi}$ yields
\be
\label{eq9}
\frac{d\lambda(\phi)}{d\phi}= \pm \frac{\kappa}{\sqrt{2}} \left(
\frac{A_S}{A_G} - \frac{A_G}{A_S} \right) \lambda .
\ee
Note that the reconstruction equation derived by HB [their Eq.\ (2.10)] has
only the first term on the right hand side of Eq.\ (\ref{eq9}), indicating
their assumption of slow-roll behavior (which here amounts to neglecting terms
of order $A_G^2/A_S^2$).

Substituting Eq.\ (\ref{eq9}) into Eq.\ (\ref{eq10}) gives
\be
\label{eq:NEWSTUF}
\frac{\lambda}{A_G(\lambda)}\, \frac{dA_G(\lambda)}{d\lambda} =
\frac{A_G^2(\lambda)}{A_S^2(\lambda)-A_G^2(\lambda)}.
\ee
This is a very important equation, because it is valid for any inflaton
potential and indicates a strong connection between the forms of the scalar
and tensor spectra produced by inflation.  The left hand side is essentially
just half of the (scale-dependent) spectral index of the tensor spectrum.
Potentially, this provides a powerful discriminator as to the correctness of
inflation. We shall refer to it as the {\em consistency equation}.
It highlights the asymmetry in the correspondence between the scalar and tensor
spectra. If one were given the tensor spectrum, then a simple differentiation
supplies the unique scalar spectrum. However, if
a scalar spectrum is supplied, then this first-order differential equation
must be solved to find the form of $A_G(\lambda)$. This leaves an undetermined
constant in the tensor spectrum and, as the consistency equation is nonlinear,
this implies that the scalar spectrum alone does not uniquely specify the
functional form of the tensors. However, knowledge of the amplitude of the
tensor spectrum at one scale is sufficient to determine this constant and lift
the degeneracy.

It is the tensor spectrum one requires to proceed with reconstruction. Once the
form of the tensor spectrum has been obtained, either directly
from observation or by integrating Eq.\ (\ref{eq:NEWSTUF}), the potential, as
parametrized by $\lambda$, may be derived by substituting Eqs.\ (\ref{eq2a})
and (\ref{eq2b}) into Eq.\ (\ref{eq1a}). This gives
\be
\label{eq13}
V[\phi(\lambda)] = \frac{16\pi^3A_G^2(\lambda)}{\kappa^4}\left[ 3 -
\frac{A_G^2(\lambda)}{A_S^2(\lambda)} \right] ,
\ee
where the final term in the square brackets again improves on HB. Finally,
integration of Eq.\ (\ref{eq9}) yields the function $\phi=\phi(\lambda)$ as
\be
\label{eq14}
\phi(\lambda) = \pm \frac{\sqrt{2}}{\kappa } \int^\lambda
\frac{d\lambda'}{\lambda'}   \frac{A_S(\lambda')A_G(\lambda')}
{A_S^2(\lambda')-A_G^2(\lambda')}
 =  \pm \frac{\sqrt{2}}{\kappa} \int^{A_G} dA'_G \frac{A_S[A'_G]}{A'^2_G} ,
\ee
where we have absorbed the integration constant by taking advantage of the
freedom to shift $\phi$ by a constant. The second integral follows after
substitution of the consistency equation and is appropriate if the functional
form of $A_S$ as a function of $A_G$ is known. The functional form of $V(\phi)$
follows by inverting Eq.\ (\ref{eq14}) and substituting the result into Eq.\
(\ref{eq13}).

The reconstruction equations are Eqs.\ (\ref{eq:NEWSTUF}), (\ref{eq13}) and
(\ref{eq14}). We emphasize again that even an arbitrarily accurate
determination of the scalar spectrum will not allow one
to determine the inflaton potential --- at least a minimal knowledge of the
tensors is required. Ultimately, though, one might hope to overdetermine the
problem by having observational knowledge of both spectra over a range of
scales. The consistency equation  (\ref{eq:NEWSTUF}) must then be satisfied, or
the inflationary hypothesis has been disproved (up to the accuracy of the
slow-roll approximation for the perturbation spectra).

The reconstruction equations allow a functional reconstruction of the inflaton
potential. For suitably simple spectra, this can be done analytically, and in
an accompanying paper \cite{CKLL} we illustrate this for the well-known cases
of scalar spectra which are exactly scale-invariant, logarithmically corrected
from scale-invariance and exact power-laws. The earliest observations with an
accuracy useful for our purposes are likely  to only provide such simple
functional fits. For advanced observations, however, one might expect that the
reconstruction equations would have to be solved numerically. There are
additional issues related to observational errors which we do not investigate
here (but see Ref.\ \cite{CKLL}).

An alternative approach, useful for obtaining mass scales, is to concentrate
on data around a given length scale $\lambda_0$, and perturbatively derive the
potential around its corresponding scalar field value $\phi_0 \equiv
\phi(\lambda_0)$. If we know $A_G(\lambda_0)$ and $A_S(\lambda_0)$ separately,
then $V(\phi_0)$ follows immediately from Eq.\ (\ref{eq13}). In order to make
further progress, one also needs information regarding the derivatives of the
spectra. Of course, the measurement of these derivatives requires knowledge of
the spectra over at least a limited range of scales, so this process is
equivalent to a Taylor expansion of the functional reconstruction \cite{MST}.

To obtain $V'(\phi)$, one needs only the derivative of the scalar spectrum, or
equivalently its spectral index. This is fortunate, as its tensor equivalent
would be much harder to observe. With the scalar spectral index $n$ (in general
a function of scale) defined as usual by
\be
1-n = \frac{{\rm d} \ln A_S^2(\lambda)}{{\rm d} \ln \lambda} ,
\ee
 one can show that
\be
\vzp  \equiv  \left. \frac{dV(\phi)}{d\phi}\right|_{\lambda=\lz} = \pm
\frac{16\pi^3}{\sqrt{2}\kappa^3} \frac{A_G^3(\lz)}{A_S(\lz)}
 \left[ 7-n_0 - (5-n_0) \frac{A_G^2(\lz)}{A_S^2(\lz)} \right] ,
\ee
where $n(\lambda_0) \equiv n_0$. If one wishes, this can be simplified into the
slow-roll approximation (in which $n_0 \approx 1$) by
ignoring the final term in the square brackets.

One can continue this process. At no stage is knowledge of the tensor spectrum
derivative required, because the consistency equation can always be used to
remove it. Given the second derivative of the scalars (equivalently the first
derivative of the scalar spectral index), one can derive an expression for
$V''(\phi_0)$, quoted in Ref.\ \cite{CKLL}, but it is too cumbersome to
reproduce here. Its slow-roll limit does not require $n_0^{\prime}$, and is
\be
\label{2DERSIM}
V_{{\rm sr}}''(\pz)  =  \frac{4\pi^3}{\kappa^2} \frac{A_G^2(\lz)} {A_S^2(\lz)}
\left[ 4(n_0 -4)^2 A_G^2(\lz)\right.
 \left. - (1-n_0)(7-n_0)A_S^2(\lz) \right].
\ee
This offers the prospect of determining whether the inflaton potential is
concave or convex when the presently observable universe crossed outside the
Hubble radius during inflation. We note immediately that $V''$ is positive if
$1<n_0<7$. It is the amplitude of the tensor perturbations at a particular
scale which yields information regarding the mass scale at which these
processes are occurring during inflation. The steepness of the potential,
measured by the dimensionful parameter $V(\phi_0)/|V'(\phi_0)|$, is determined
by the ratio $A_S(\lambda_0)/A_G(\lambda_0)$.

Let us illustrate by example. Within a few years a combination of microwave
background anisotropy measurements should give us some information about the
scalar and tensor amplitudes at a particular length scale $\lambda_0$
(corresponding to an angular scale $\theta_0$) \cite{CBDES}. A hypothetical,
but plausible, data set that this might provide would be $A_S(\lambda_0) = 1
\times 10^{-5}$; $A_G(\lambda_0) = 2 \times 10^{-6}$; $n_0 = 0.9$. This would
lead to
\bea
V(\phi_0)   & = &      ( 2 \times 10^{16} {\rm GeV} )^4 \nonumber \\
\pm V'(\phi_0)  & = &   ( 3 \times 10^{15} {\rm GeV} )^3 \nonumber \\
V_{{\rm sr}}''(\phi_0) & = &      ( 5 \times 10^{13} {\rm GeV} )^2.
\eea
In this way cosmology might be first to get a ``piece of the action'' of
GUT--scale physics.

In this {\em Letter} we have discussed the promising possibility of large
scale structure observations, particularly of tensor modes, providing rather
specific information as regards the physics of the Grand Unified era. We have
derived equations which allow a knowledge of either the scalar spectrum, the
tensor spectrum, or preferably both, to be used to reconstruct the potential of
 the inflaton field. We have also noted a consistency equation, by which the
scale-dependences of the spectra {\em must} be related if their origin lies in
an inflationary era. This potentially provides a powerful test of inflation;
the minimum knowledge required to implement it would be knowledge of the scalar
spectrum across a range of scales plus the amplitude of the tensor spectrum at
{\em two} of the wavelengths. [Technically the minimum is the tensor spectrum
plus the scalar amplitude at a single scale, but observationally that would be
considerably more demanding.]

In an accompanying paper \cite{CKLL}, all the issues herein are discussed in
greater detail. As well as providing examples of functional reconstruction, we
discuss in detail the opportunities available in both presently available and
expected future observations for carrying out the program we have outlined
here. While the ambitious aim of full reconstruction appears to lie some way
into the future, we are optimistic as to the short-term possibilities, as
tantalizingly indicated in \cite{CBDES}, of obtaining at least a perturbative
reconstruction of the inflaton potential and a window on GUT scale physics.

EJC, ARL and JEL are supported by the Science and Engineering Research Council
(SERC) UK. EWK and JEL are supported at Fermilab by the DOE and NASA under
Grant NAGW--2381. ARL acknowledges the use of the Starlink computer system at
the University of Sussex.  We would like to thank D.\ Lyth, P.\ J.\
Steinhardt, and M.\ S.\ Turner for helpful discussions.
%%%%%%%%%%%%%%%%%%%%%%%%%%%%%%%%%%%%%%%%%%%%%%%%%%%%%%%%%%%%%%%%%%%%%%
\frenchspacing
\def\prl#1#2#3{{\em Phys. Rev. Lett.} {\bf #1}, #2 (#3)}
\def\prd#1#2#3{{\em Phys. Rev. D} {\bf #1}, #2 (#3)}
\def\plb#1#2#3{{\em Phys. Lett.} {\bf #1B}, #2 (#3)}
\def\npb#1#2#3{{\em Nucl. Phys.} {\bf B#1}, #2 (#3)}
\def\apj#1#2#3{{\em Astrophys. J.} {\bf #1}, #2 (#3)}
\def\apjl#1#2#3{{\em Astrophys. J. Lett.} {\bf #1}, #2 (#3)}
%%%%%%%%%%%%%%%%%%%%%%%%%%%%%%%%%%%%%%%%%%%%%%%%%%%%%%%%%%%%%%%%%%%%%

%%%%%%%%%%%%%%%%%%%%%%%%%%%%%%%%%%%%%%%%%%%%%%%%%%%%%%%%%%%%%%%%%%%%%%
\end{document}